\def\gr{$\gamma$-ray }
\def\grs{$\gamma$-rays }
\def\be{\begin{equation}}
\def\ee{\end{equation}}
\def\part#1#2{{{\partial #1}\over {\partial #2}}}

\documentclass{aa}
\usepackage{times}
\usepackage{graphics}

\begin{document}

\title{Channeled blast wave behavior based on longitudinal instabilities}
\author{M. Pohl\inst{1} \and I. Lerche\inst{2} \and R. Schlickeiser\inst{1}}
\institute{Institut f\"ur Theoretische Physik,
Lehrstuhl IV: Weltraum- und Astrophysik, Ruhr-Universit\"at Bochum,
D-44780 Bochum, Germany \and
Department of Geological Sciences, University of South Carolina,
Columbia, SC 29208, USA}
\date{Received ; accepted }
\offprints{M. Pohl, \email{mkp@tp4.ruhr-uni-bochum.de}}
\titlerunning{Longitudinal instabilities}

\abstract{ To address the important issue of how kinetic energy of
collimated blast waves is converted into radiation, Pohl \& Schlickeiser
(\cite{ps00}) have recently investigated the relativistic two-stream instability
of electromagnetic turbulence. They have shown that swept-up 
matter is quickly iso\-tro\-pized in the blast wave, which provides 
relativistic particles and, as a result, radiation. 
Here we present new calculations for the electrostatic instability in such
systems. It is shown that the electrostatic instability is faster than 
the electromagnetic instability for highly relativistic beams. However,
even after relaxation of the beam via the faster electrostatic turbulence,
the beam is still unstable with respect to the electromagnetic waves,
thus providing the isotropization required for efficient production
of radiation. While the emission spectra in the model of 
Pohl and Schlickeiser have to be modified, the basic characteristics
persist.
}

\maketitle

\section{Introduction}
Published models for the \gr emission of blazars are usually based
on interactions of highly relativistic electrons of unspecified origin
propagating in the relativistic jet of active galactic nuclei
(Hartman et al. \cite{ha97}, Bicknell et al. \cite{bi00}, Catanese
and Weekes \cite{cw99}). The usual shock
or stochastic electron acceleration processes would have to be very fast to
compete efficiently with the strong radiative losses at high electron energies,
{which operate on timescales of the order of ten seconds for the 
rapidly variable \gr blazars, which is similar to the duration of a single 
shock crossing cycle for TeV electrons at a moderately relativistic shock
assuming Bohm diffusion (Gallant \& Achterberg \cite{ga98}).}
Energetic electrons may also result as secondaries from photomeson production of
highly relativistic hadrons, but the observed short variability timescale places
extreme conditions on the magnetic field strength, for the hadron gyroradius has
to be much smaller than the system size, and on the Doppler factor, for the
intrinsic timescale for switching off the cascade is linked to the observed soft
photon flux.
\begin{figure}
\resizebox{\hsize}{!}{\includegraphics{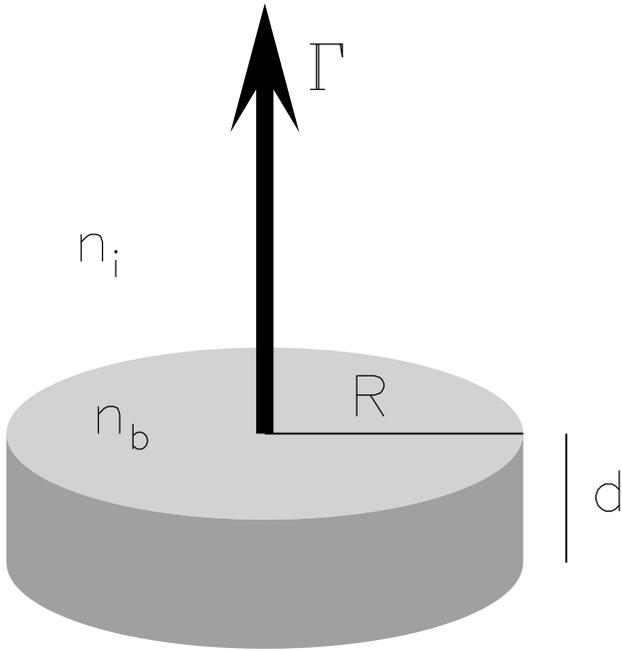}}
\caption{Sketch of the basic geometry. The thickness of the
channeled blast wave $d$, measured in its rest frame, is much smaller than its halfdiameter. The blast wave
moves with a bulk Lorentz factor $\Gamma$ through ambient matter
of density $n_{\rm i}$.}
\label{sketch}
\end{figure}

Here we consider the energisation of relativistic particles in jets by
interactions with the surrounding medium. {The rapid variability displayed 
in the \gr light curves of blazars requires the emission regions be less than 
$\sim\,$0.01 pc in size. Also, VLBI observations of blazars 
indicate that the jets are not continuously filled
emission regions, but consist of individual structures which 
relativistically move along a common trajectory and can be followed 
over years. On larger scales optical data still show individual knots, but also emission in-between (e.g. Boksenberg et al. \cite{bok92}). Detailed spectral 
studies (e.g. Meisenheimer, R\"oser and Schl\"otelburg \cite{mrs96})
indicate that the spectrum of the synchrotron radiating electrons is 
fairly independent of position, suggesting that many acceleration sites 
with uniform characteristics are operational on scales of $\sim\,$100 pc.

Accounting for these findings we model a jet as
a channeled outflow with relativistic 
bulk velocity $V$, consisting of isolated plasma clouds or blast waves
which contain
cold electrons and protons of density $n_b$ (see Fig.\ref{sketch}). 
These clouds may correspond to the individual components observed in 
VLBI images of blazars, whereas many of them and secondary electrons of escaped
nucleons would contribute to the optical 
and X-ray jets observed on larger scales in nearby AGN such as M87.
For convenience we assume that the outflow
is directed parallel to the uniform background magnetic field. We also
neglect a possible expansion of the jet cloud in this paper. Expansion is 
expected to occur on much longer timescales than the effects discussed here. }

Each cloud of protons and electrons 
propagates into the surrounding interstellar medium that consists
of cold protons and electrons of density $n_i^*$ (quantities indexed
with an asterisk are measured in the laboratory (galaxy) frame). 
Viewed from the coordinate system
comoving with the outflow, the interstellar protons and electrons represent a
proton-electron beam propagating with relativistic speed $-V$ 
antiparallel to the
uniform magnetic field direction. 
This situation is unstable
and waves will be excited which backreact on the incoming beam. In our
earlier analysis (Pohl \& Schlickeiser \cite{ps00})
the stability of this beam was examined under the assumption
that the background magnetic field is uniform and directed parallel to the 
direction of motion. It was shown that the beam very quickly excites
low-frequency electromagnetic waves, which quasi-linearly isotropize the
incoming interstellar electrons and protons in the blast wave plasma, thus
providing relativistic particles. {These processes are essentially those thought
to operate in the first half-cycle of relativistic shock acceleration,
but rather than assuming the existence of a hydrodynamical shock and a 
particular intensity of MHD turbulence, kinetic theory was used to 
self-consistently calculate the build-up of turbulence and its 
backreaction on the incoming interstellar particles.

The protons carry the bulk of the power,
because velocities are conserved in the isotropization process and hence the
protons have a factor of $\sim\,$2000 more kinetic energy than the electrons.
The protons produce high energy emission via the production of neutral
pions in inelastic pp collisions. 
In parallel charged pions would provide secondary electrons which emit
synchrotron radiation as well as bremsstrahlung and inverse Compton scattering.
As discussed in Pohl \& Schlickeiser (\cite{ps00}), }
the high energy emission produced by these
particles has characteristics typical of BL Lacertae objects. 
The observed secular variability of the \gr emission of AGN would be
related to the existence (or non-existence)
of a relativistic blast wave in the sources, and thus to the availability
of free energy in the system.  
The observed fast variability, on the other hand, would
be caused by density inhomogeneities in the interstellar medium
through which the blast wave propagates.

In parallel
to the \grs, neutrinos are emitted whose spectrum and flux would be closely
correlated with those of the $\gamma$-rays, which permits one to use the \gr
light curves of blazars to very efficiently search for neutrino emission
as a diagnostic for an hadronic origin of the high energy radiation
(Schuster et al. \cite{sps01}).

{In the initial stage virtually no radio emission would be produced, for the
plasma frequency and the synchrotron self-absorption frequency would be too
high. After the initial \gr emission phase, 
when the blast wave has decelerated and expanded, a mm-radio flare
would build up, which behaviour is preferentially observed in radio
to \gr correlation studies of EGRET sources (M\"ucke et al. \cite{mue96}).}

Here we expand on the previous
treatment by calculating the two-stream instability for
longitudinal, electrostatic waves. 
In contrast to electromagnetic waves,
which scatter the particles in pitch angle but preserve their kinetic energy
until the distribution is isotropized,
the electrostatic waves change the particles' energy until a 
plateau distribution is established.

\section{Longitudinal instabilities of a proton-electron beam}
\subsection{The dispersion relation}

For charge $e$, mass $m$ particles under the action of an
electrical field $\vec E = \vec e_\parallel E_\parallel \,\exp(\imath k(x-at))$,
in the direction parallel to an ambient magnetic field the perturbations,
$\delta\! f$, to the original distribution function satisfy
\be
\part{\delta\! f}{t} + \vec v \part{\delta\! f}{\vec x} + e\vec E \part{f}{\vec p}=0
\label{9}
\ee
which, to first order in $\delta\! f /f_0$, and with $\delta\! f$ also
taken to vary as
$\delta\! f(p) \exp(\imath k(x-at))$ gives
\be
\imath k (v_\parallel - a) \delta\! f + e E_\parallel \part{f_0}{p_\parallel}=0
\label{10}\ee
\be {\rm i.e.}\qquad\qquad \delta\! f = {{\imath e E_\parallel}\over {k
(v_\parallel - a)}} \part{f_0}{p_\parallel}
\label{11}\ee
The electrostatic balance equation requires
\be {\rm div} \vec{E} = 4\pi \sum  e\int \delta\! f\,d^3p\label{12}\ee
where the sum is over charge species. Thus
\be
k^2 E_\parallel = 4\pi\,E_\parallel\sum e^2 \int {{\part{f_0}{p_\parallel}}\over
{v_\parallel -a}}\,d^3p
\label{13}\ee
Then the dispersion relation is 
\be k^2 = 4\pi \sum e^2 \int {{\part{f_0}{p_\parallel}}\over
{v_\parallel -a}}\,d^3p
\label{14}\ee
Let $a= a_{\rm R} + \imath a_{\rm I}$. {Then split the RHS of eq.~\ref{14} formally into the real part, $J$, and the imaginary part, $I$,}
\be k^2 = J(a_{\rm R},a_{\rm I}) + \imath I(a_{\rm R},a_{\rm I}) \label{15}\ee
Because $k^2$ is meromorphic, then
\be
I(a_{\rm R},a_{\rm I})=0 \simeq I(a_{\rm R},0)+ a_{\rm I} \part{I(a_{\rm R},a_{\rm I})}{a_{\rm I}}\Big\vert_{a_{\rm I}=0}
\label{16}\ee
The Cauchy condition yields
\be\part{I(a_{\rm R},a_{\rm I})}{a_{\rm I}}=\part{J(a_{\rm R},a_{\rm I})}{a_{\rm R}}\label{17}\ee
Thus the lowest order approximation yields
\be a_{\rm I} \simeq - I(a_{\rm R},0)\,\left(\part{J(a_{\rm R},0)}{a_{\rm R}}\right)^{-1}\label{18}\ee
Then 
\be
J(a_{\rm R},0) = 4\pi \sum e^2 {\rm P}\int {{\part{f_0}{p_\parallel}}\over
{v_\parallel -a_{\rm R}}}\,d^3p
\label{19}\ee
{where $P$ denotes the Cauchy principal value}, while 
\be I(a_{\rm R},0) = 4\pi^2\sum e^2 \int \part{f_0}{p_\parallel}\,\delta (v_\parallel
-a_{\rm R})\,d^3p\label{20}\ee
Using the notation of Pohl \& Schlickeiser (\cite{ps00}) the initial distribution function in the blast wave frame is (after correction of a typo
in their Eq.3)
\be
f(\vec{p},t=0)=
{{n_{\rm i}\delta \bigl(
p_{\perp }\bigr)\delta \bigl(p_{\parallel }+P\bigr)}\over {2\pi p_{\perp }}}+\; 
 {{n_{\rm b}\delta \bigl (p_{\perp }\bigr)
\delta \bigl(p_{\parallel }\bigr)}\over {2\pi p_{\perp }}}
\ ,\label{dist_f}\ee
In this case Eq.~\ref{14} reduces to
\be
k^2 = \Omega^2\left[{1\over {a^2}} + 
{{\delta}\over {(a + V)^2}}\right]\label{21}\ee
where $\Omega^2 = \omega_{p,p}^2 + \omega_{p,e}^2$ and
$\delta=n_i/ \left(n_b\,\Gamma^3\right)$.

\subsection{Properties of the dispersion relation}

\begin{figure}
\resizebox{\hsize}{!}{\includegraphics{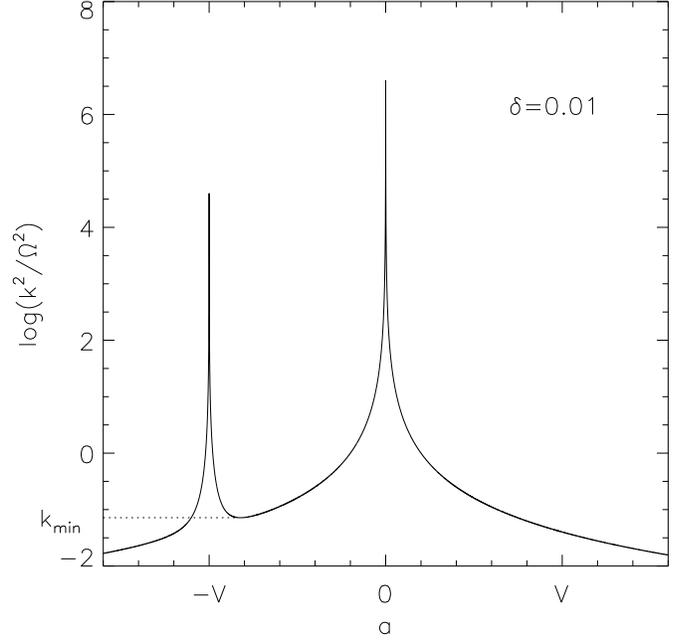}}
\caption{The dispersion relation for longitudinal waves. Note that there is always a real solution in the range $a\le -V$, and another real solution in 
$a\ge 0$. In the remaining range $-V\le a\le 0$ the dispersion relation returns
two real values of $a$ only if $k^2$ exceeds $k_{\rm min}^2$.}
\label{fig2}
\end{figure}

A plot of $k^2$ versus $a$ is given in Fig.~\ref{fig2}. To be noted from the figure is that there are always two real values of $a$, occurring in $a> 0$
and $a< -V$, respectively. In the region $-V \le a\le 0$ the dispersion relation returns two real positive values of $a$, provided $k^2$ exceeds a minimum value,
$k_{\rm min}^2$, given by
\be
k_{\rm min}^2 = {{\Omega^2}\over {V^2}} (1+\delta^{1/3})^3
\label{23}\ee
occurring on
\be
a_{\rm min} =  -V (1+\delta^{1/3})^{-1} \ > - V
\label{24}\ee

\begin{figure}
\resizebox{\hsize}{!}{\includegraphics{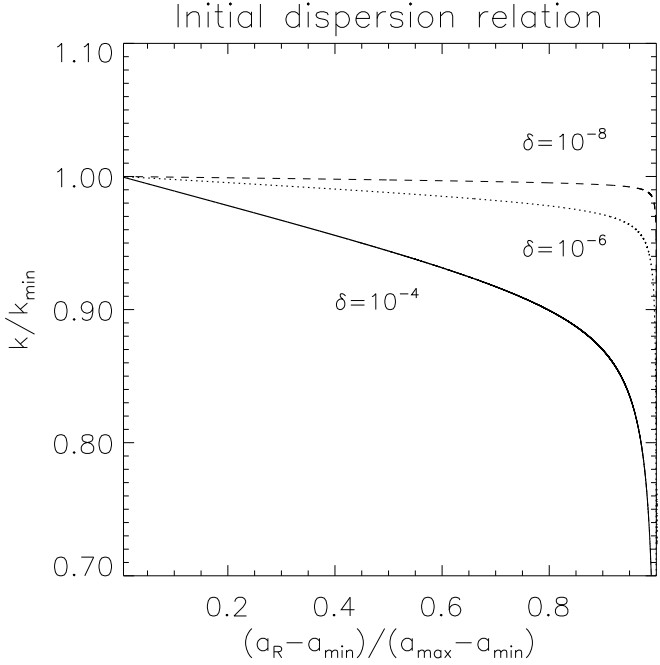}}
\caption{An enlargement of the dispersion relation for longitudinal waves
in the phase velocity range where instability occurs. For small perturbation
parameters, $\delta$, $k/k_{\rm min}$ is essentially flat, unless 
$a_{\rm R}$ is very close to $a_{\rm max}$.}
\label{k-a}
\end{figure}

For values $k^2 < k_{\rm min}^2$, the solutions are complex conjugate 
solutions in $-V\le a_{\rm R}\le 0$. Note that on $k=0$, the complex solutions are
\be a_{\rm 0} = - V (1+\delta)^{-1} (1\pm \imath \sqrt{\delta})
= a_{\rm max}\,(1\pm \imath \sqrt{\delta})\label{26}\ee
For more general values of $k^2 < k_{\rm min}^2$, the general solution
to the dispersion relation is given through
\be
a\,V^{-1} =  - \sin^2 \phi + \imath y_{\rm I}\label{27}\ee
with 
\be
{{k^2}\over {k_{\rm min}^2}}  = 
\left( {{\tan^2\phi +1}\over {\tan^2\phi -1}}
\right)^2 {{(\tan\phi - \sqrt{\delta})(1-\tan\phi \sqrt{\delta})}\over
{\tan\phi\, (1+\delta^{1/3})^3}}\label{28}\ee
and 
\be
y_{\rm I}^2 = {{\tan\phi (\sqrt{\delta} \tan^3\phi -1)}\over
{(1+\tan^2\phi)^2 (\tan\phi - \sqrt{\delta})}}\label{30}\ee
The range of $\phi$ is restricted to
\be \delta^{-1/6} \le \tan\phi\le \delta^{-1/2}\label{31}\ee
On $\tan\phi = \delta^{-1/6}$ note that $k^2=k_{\rm min}^2$ and 
$a_{\rm R} =a_{\rm min}$, and
on $\tan\phi = \delta^{-1/2}$, $k=0$ and 
$a_{\rm R} =a_{\rm max}$. Thus the range (\ref{31}) covers the 
complete wavenumber spectrum where instability can occur. In Fig.~\ref{k-a}
we show an enlargement of the dispersion relation
in the phase velocity range where instability occurs. It is obvious
that $k^2$ is basically independent of the phase velocity $a_{\rm R}$, 
unless $a_{\rm R}$ is very close to $a_{\rm max}$. Therefore the Taylor
expansion to first order (Eq.~\ref{18}) is a poor approximation,
for $\part{J(a_{\rm R},0)}{a_{\rm R}} \simeq 0$. A second order expansion would provide
us only with $a_{\rm R}^2$ which we have already derived in Eq.~\ref{30}.
However, Eq.~\ref{18} is very useful for determining the sign of the growth 
rate. Noting that $a_{\rm R} < 0$, we see that $I(a_{\rm R},0)$ is negative for 
negative $\part{f_0}{p_\parallel}$ because $\part{v_\parallel}{p_\parallel}
> 0$. $J(a_{\rm R},0)$ is given in Eq.~\ref{21}, provided $a$ is replaced by
$a_{\rm R}$. Obviously 
$\part{J(a_{\rm R},0)}{a_{\rm R}} < 0$, and thus $a_{\rm I} < 0$. 
Since $k<0$ we get for the growth rate
\be
\gamma = k a_{\rm I} = \vert k\vert\,\vert a_{\rm I}\vert > 0
\label{gro}\ee
As in the non-relativistic case parallel electrostatic waves grow 
as long as the distribution function is inverted, {whereas
relativistic Landau damping would occur for $\part{f_0}{p_\parallel} > 0$ 
with $p_\parallel < 0$.}

\begin{figure}
\resizebox{\hsize}{!}{\includegraphics{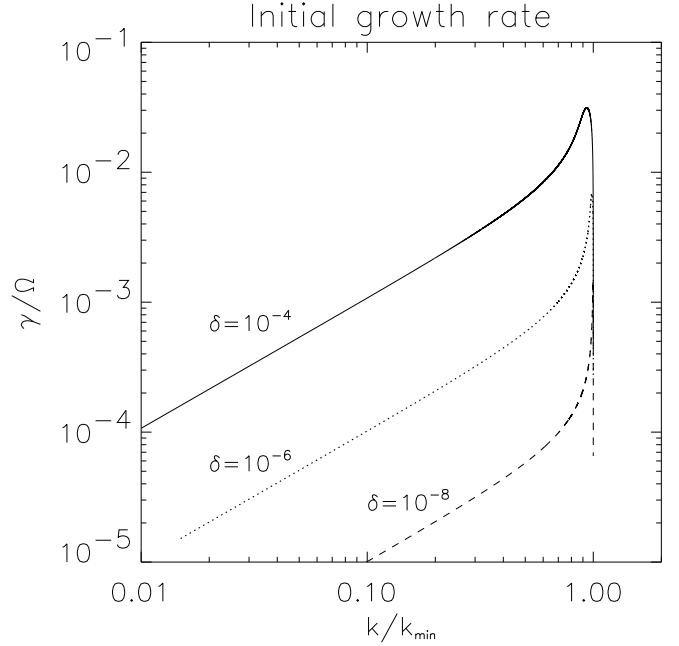}}
\caption{The growth rate $\gamma$ as a function of the wave number $k$ for 
three values of the perturbation parameter $\delta=n_i/ 
\left(n_b\,\Gamma^3\right)$. Most of the instability occurs in a small
wavenumber range close to $k_{\rm min}$.}
\label{growth_k}
\end{figure}

\begin{figure}
\resizebox{\hsize}{!}{\includegraphics{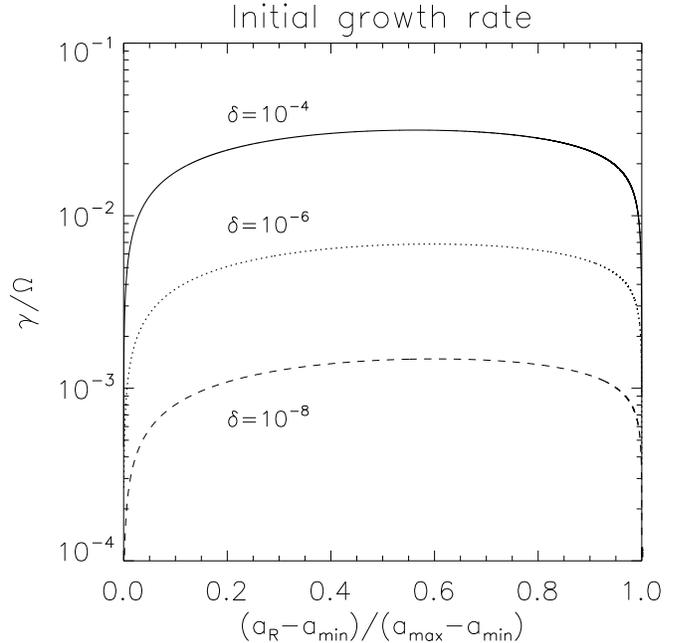}}
\caption{The growth rate $\gamma$ as a function of the phase velocity 
$a_{\rm R}$ for three values of the perturbation parameter $\delta$.
The perturbation parameter controls only the speed of the instability, not
its spectral form.}
\label{growth_a}
\end{figure}

The square of the growth rate is
\be 
\gamma^2 = k^2 V^2 y_{\rm I}^2 = \Omega^2 {{(\sqrt{\delta} \tan^3\phi -1)
(1-\tan\phi \sqrt{\delta})}\over
{(\tan^2\phi -1)^2}}\label{32}\ee
In Figs.~\ref{growth_k} and \ref{growth_a} we show the growth rate
as a function of the wavenumber $k$ and phase velocity, $a_{\rm R}$,
respectively. The growth rate is sharply peaked at wavenumbers close to
$k_{\rm min}$, but is essentially independent of the phase velocity, $a_{\rm R}$.


\subsection{Quasi-linear behavior}

The time-dependent behavior of the intensities $I(k,t)$ of the
excited waves is given by (Lerche \cite{le67}, Lee \& Ip \cite{li87}) 
\be \part{ }{t} I(k,t) = 2\, \gamma\, I(k,t)\label{36}\ee
Note that instability occurs only for negative wave phase speeds $a_{\rm R}$, implying
negative wavenumbers $k$.

To describe the long-term influence of the excited waves on the beam particles,
which are the particles that can resonate with the waves, one uses the 
quasi-linear Fokker-Planck equation.
Because the longitudinal waves act with an electric vector only, and because 
that vector parallels the magnetic field, the phase space density for the
resonant particles then has only its momentum parallel to the ambient field 
influenced by the longitudinal turbulence. Hence, the corresponding
Fokker-Planck equation reads
\be \part{f}{t} = \part{ }{p_\parallel} \left( D \part{f}{p_\parallel}\right)
\label{37}\ee
where the diffusion coefficient, $D$, is given by
\begin{displaymath}
D = {{\langle \Delta p_\parallel^2\rangle}\over {\Delta t}}
= 16 \pi^2 e^2 \int_{k_{\rm min}}^0 dk \ I(k)\,
\delta\left[k(v_\parallel -a_{\rm R})\right]
\end{displaymath}
\be
\hphantom{D} = 16 \pi^2 e^2 \int_{k_{\rm min}}^0 dk \ 
{{I(k)}\over {\vert k\vert}}\,
\delta(v_\parallel -a_{\rm R})
\label{38}\ee
The corresponding diffusion equation for each resonant particle species is
\be
\part{f}{t} = 16 \pi^2 e^2 \part{ }{p_\parallel} \left( 
\int_{k_{\rm min}}^0 dk \ 
{{I(k)}\over {\vert k\vert}}\,
\delta(v_\parallel -a_{\rm R})\,\part{f}{p_\parallel}\right)
\label{40}\ee
where $a_{\rm R}$ is related to the wavenumber $k$ through equation (\ref{28}).

Note that some care has to be exercised in treating the $\delta$-function
in equation (\ref{40}) for two reasons. First, slight
variations in $a_{\rm R}$ ($v_\parallel$) correspond to large variations in 
$p_\parallel$ for relativistic particles. Second, during the evolution
of the system the initial beam may disperse, for example through interactions
with electromagnetic turbulence, so that a range of $p_\parallel$
would correspond to a given $v_\parallel$.
We consider here two extreme cases:

(1) when the variation in phase speed is much larger than the variation
in transverse speed;

(2) when the variation in transverse speed is much larger than the variation
in phase speed.

Consider each case in turn.

\subsection{Highly variable phase speed}

In this situation one can effectively set the resonant particles to a 
cold gas approximation and deal only with the bulk flow. 

One conventionally argues that the wave generation is sufficiently rapid
that the local (in time) wave intensity has reached a quasi-steady state
in the sense that it alters quasi-adiabatically as the particles change
their bulk streaming parameters to feed the waves. 

Because of the unknown nature of the initial wave spectrum, and because other
processes ignored in the development will also influence the generation
of waves, one of the standard devices is to ignore the wave intensity spectrum
generation and to use {\it models} of how one believes the waves spectrum
has evolved to its current state.
This particular decoupling of the wave and particle behaviours has been an 
enormous success in cosmic ray astrophysics in general and in heliospheric
physics (A thorough review can be found in Schlickeiser, \cite{sch00}).
Under this scenario one writes equation (\ref{38}) now with $I(k,t)$
a wave intensity spectrum to be prescribed by the user.

In addition, the parameters related to the relaxation of the plasma,$V$, 
$a_{\rm R}$, etc., are considered to be independent of time in evaluating
the diffusion coefficient (\ref{38}), but are then allowed to evolve
with time once the diffusion coefficient is evaluated in order that
the long term evolution of the plasma distribution function can be tracked.

The particle velocities, $v_\parallel$, for which resonance can be 
established, are those which satisfy
\be 
{{-V}\over {1+\delta}} \le v_\parallel \le {{-V}\over {1+\delta^{1/3}}}
\label{61}\ee
corresponding to particle momenta, $p_\parallel$, in the range
\be
{{- V \Gamma}\over \sqrt{1+\delta\Gamma^2 (2+\delta)}} \le
{p_\parallel\over m} \le 
{{-V\Gamma}\over \sqrt{1+\delta^{1/3}\Gamma^2 (2+\delta^{1/3})}}
\label{62}\ee
Clearly, the range of resonant particle momenta is sensitive to
$2\delta \Gamma^2$ and $2\delta^{1/3} \Gamma^2$. If both are small 
compared to unity then the range of resonant momenta is
close to the initial momentum of the beam particles, $-P$, and
has a width of $\sim P \delta^{1/3} \Gamma^2$, while,
if $2\delta \Gamma^2$ and $2\delta^{1/3} \Gamma^2$ are both large 
compared to unity then the particle momentum range is far from $-P$
and has a width of $\sim P/\sqrt{2\delta \Gamma^2}$. Thus the
quasi-linear evolution is sensitive to the values of
$2\delta \Gamma^2$ and $2\delta^{1/3} \Gamma^2$, both of which are
required to be small compared to unity in the astrophysical application to AGN 
(Pohl \& Schlickeiser \cite{ps00}). 

When we consider the growth rate (Eq.\ref{32}) as a function of
phase velocity, we see in Fig.~\ref{growth_a} that
the growth rate is essentially independent of $a_{\rm R}$. Neglecting
the possible effects of damping and cascading, we can therefore expect
that the intensity spectrum in phase velocity, $I(a_{\rm R}) = I(k)
\part{k}{a_{\rm R}}$ is flat between $a_{\rm min}$ and $a_{\rm max}$
and zero outside this range (Akhiezer et al. \cite{ak75}). Upon
interaction with this wave spectrum the beam will relax to a plateau
distribution between the velocities $-V$ and $v_\parallel = a_{\rm min}$. 
The energy available for the build-up of the waves comes from the beam.
Because the wave growth is much faster than the relaxation of the beam,
we may estimate the final wave intensity spectrum as the ratio of the
energy lost by the beam during relaxation to a plateau 
between the velocities $-V$ and $v_\parallel =  a_{\rm min}$ and the 
phase velocity range in which effective growth occurs, namely
$\delta^{1/3} V $.
Therefore we may write
\be
\Delta E \simeq {{n_i}\over 2}\, m_p c^2\,\Gamma^3\,\delta^{1/3}\ \label{55a}
\ee
\be
\Delta a_{\rm R} \simeq V\,\delta^{1/3}\ \label{55b}\ee
\be
I(a_{\rm R}) \simeq {{\Delta E}\over {\Delta a_{\rm R}}} \simeq
{{n_i\,m_pc^2\,\Gamma^3}\over {2\,V}} = I_0\ \label{55c}\ee
where we keep in mind that $a_{\rm R}$ is to be taken between $a_{\rm min}$
and $a_{\rm max}$. 
The diffusion coefficient is then
\begin{displaymath}
D(p_\parallel ) = 16\pi^2 e^2\,I_0 
\int_{a_{\rm max}}^{a_{\rm min}} da_{\rm R}\ {1\over {\vert k\vert}} 
\,\delta \left(v_\parallel -a_{\rm R}\right)
\end{displaymath}
\begin{displaymath}
\hphantom{D(p_\parallel )} = 16\pi^2 e^2\,{{I_0}\over {\vert k(a_{\rm R}=
v_\parallel)\vert}} 
\end{displaymath}
\be
\hphantom{D(p_\parallel )} \simeq
2\pi\,m_e\,m_p\,c^2\Gamma^3\,\Omega \,{{n_i}\over {n_b}}
 \label{58}\ee
where we have used again that $k\simeq k_{\rm min}\simeq
-V/\Omega$ in the resonant range.
The time scale for the relaxation of the beam by the electrostatic instability 
can then be estimated as
\be
\tau \simeq {{P^2}\over D}\simeq {{m_p\,n_b}\over {2\pi\,m_e\,n_i\,\Omega\,\Gamma}}
\simeq 5\cdot 10^{-3}\,{\sqrt{n_b}\over {\Gamma\,n_i}}\ \ {\rm sec}\ \label{64}
\ee
In term of the interstellar matter density in the laboratory frame, $n_i^\ast$,
and the scaled variables $n_8= 10^{-8} n_b$ and $\Gamma_2 =10^{-2} \Gamma$
the relaxation time scale reads
\be 
\tau =5\cdot 10^{-3}\,{\sqrt{n_8}\over {\Gamma_2^2\,n_i^\ast}}\quad {\rm sec}\ \label{65}
\ee
which is two orders of magnitude shorter than the time scale for the 
electromagnetic instability derived in Pohl \& Schlickeiser (\cite{ps00}).

\subsection{Highly variable transverse speed}

Suppose the initial beam is slightly dispersed, such that all particles
have the same total momentum, $P$, but instead of flying exactly backwards
($\mu =-1$) the particles homogeneously occupy a fixed solid angle element
defined by $\mu \le \mu_c$. This initial condition approximates the 
situation after the onset of pitch angle scattering but long before the isotropization is completed. The initial distribution function in the 
blast wave frame is then
\begin{displaymath}
f(\vec{p},t=0)={{n_{\rm b}\delta \bigl (p_{\perp }\bigr)
\delta \bigl(p_{\parallel }\bigr)}\over {2\pi p_{\perp }}} 
\end{displaymath}
\be
\hphantom{f(\vec{p},t=0)=}
+{{n_{\rm i}\delta \bigl(\sqrt{p_\perp^2 + p_\parallel^2} -P\bigr)\,
\Theta \bigl(\mu_c -{p_{\parallel}\over P}\bigr)}\over 
{2\pi\,P^2\,(1+\mu_c)}} 
\ \label{70}\ee
The dispersion relation (\ref{14}) becomes
\be
k^2 = \Omega^2\left[{1\over {a^2}} + {{\delta\,\Gamma^2}\over
{c^2\,(1+\mu_c)}}\,\int_{-1}^{\mu_c} dy\ {{\beta^{-2} - y^2}\over 
{\left(y-{a\over V}\right)^2}}
\right]\ \label{71}\ee
where $\beta = V/c$, $y=p_\parallel/P$,
and $\delta=n_i/ \left(n_b\,\Gamma^3\right)$ as in Eq.(\ref{26}). 
The integral in Eq.(\ref{71})
can be trivially solved and we obtain the dispersion relation for a
dispersed beam with $x=a/V$ and $\epsilon=1+\mu_c$
\be
{{k^2\,V^2}\over {\Omega^2}} = {1\over {x^2}}\, +\, \delta\,\Gamma^2
\,g(x)\ \label{72}\ee
with
\begin{displaymath}
g(x)={1\over \epsilon} \int_{-1}^{\epsilon -1} dy\ {{\beta^{-2} - y^2}\over 
{\left(y-{a\over V}\right)^2}}
\end{displaymath}
which for $\epsilon\rightarrow 0$ would reduce to Eq.~\ref{21}.
\begin{figure}
\resizebox{\hsize}{!}{\includegraphics{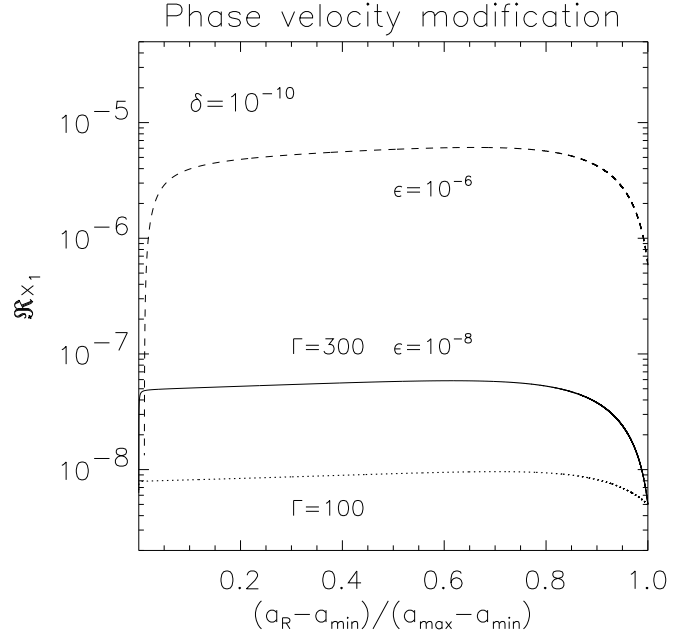}}
\caption{The normalized phase velocity modification, $\Re x_1$, for different beam openings, $\epsilon$, and different bulk Lorentz factors, $\Gamma$. The perturbation 
parameter $\delta$ is held constant. The phase velocity modification scales
approximately linearly with $\epsilon$, unless $a_{\rm R}$ 
is close to $a_{\rm min}$.
}
\label{correct_a}
\end{figure}
\begin{figure}
\resizebox{\hsize}{!}{\includegraphics{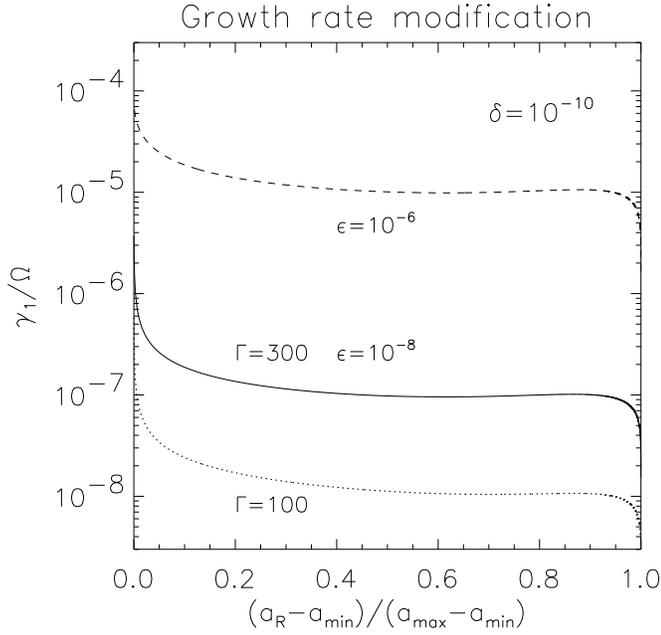}}
\caption{The growth rate modification, $\gamma_1$, as a function of the 
unmodified phase velocity 
$a_{\rm R}$ for different beam openings, $\epsilon$, and 
different bulk Lorentz factors, $\Gamma$.
The perturbation 
parameter $\delta$ is held constant. The growth rate modification has a 
similar dependence on the parameters as has the phase velocity modification.}
\label{correct_y}
\end{figure}
\begin{figure}
\resizebox{\hsize}{!}{\includegraphics{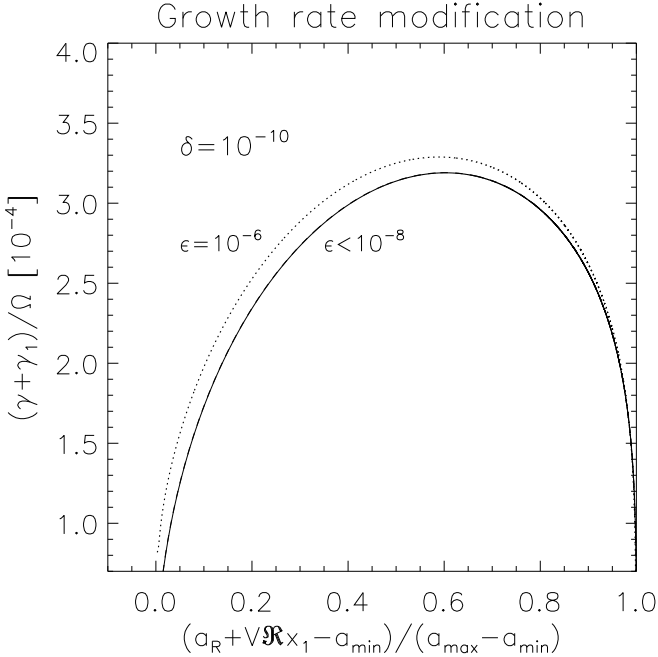}}
\caption{The modified growth rate, $\gamma+\gamma_1$, as a function of the
modified phase velocity, $a_{\rm R}+V\Re x_1$, for different beam openings, $\epsilon$. The perturbation 
parameter $\delta$ and the bulk Lorentz factor $\Gamma =300$ are 
held constant. Note that this plot has a linear scaling in units of
$10^{-4}$. The solid line applies for all $\epsilon \le 10^{-8}$, 
for which the modification has no visible effect.
The dotted line represents the enhanced growth rate for $\epsilon=10^{-6}$. 
}
\label{corr_growth}
\end{figure}
For small $\epsilon$ we can derive modifications 
\be
x=x_0 + x_1 \label{73}
\ee
to the cold beam solutions $x_0$
(Eq.\ref{27}) of the dispersion relation by expanding the right-hand-side
of Eq.\ref{72},
${\rm RHS} (x,\epsilon)$.
\begin{displaymath}
{{k^2\,V^2}\over {\Omega^2}} =  \epsilon \part{{\rm RHS}(x_0,\epsilon)}{\epsilon}
\bigg\vert_{\epsilon =0} +
x_1 \part{{\rm RHS}(x,0)}{x}
\bigg\vert_{x =x_0} 
\end{displaymath}
\be
\hphantom{{{k^2\,V^2}\over {\Omega^2}} =}+
{{x_1^2}\over 2} \part{^2 {\rm RHS}(x,0)}{x^2}
\bigg\vert_{x =x_0} + {\rm RHS}(x_0,0)\label{75}\ee
The second order expansion in $x$ is necessary because the first order term in
$x$ is zero for $\tan \phi =\delta^{-1/6}$.

The term ${\rm RHS}(x_0,0)$ is identical to the right-hand-side of the 
dispersion relation for an ideal beam (\ref{21}), and hence equal to 
$k^2 V^2/\Omega^2$. The remaining binomial expression in $x_1$ is solved by
\be
x_1=(1+x_0) {{(1+x_0^{-1})^3 + \delta - \sqrt{f(x_0,\epsilon)}}\over
{3 \left[
(1+x_0^{-1})^4 + \delta\right]}}\label{76}\ee
where 
\begin{displaymath} 
f(x_0,\epsilon)=\left[(1+x_0^{-1})^3 + \delta\right]^2
\end{displaymath}
\be 
\hphantom{f(x_0,\epsilon)=}
- 3\epsilon\delta\Gamma^2 {{1+\beta^2 x_0}\over {1+x_0}}\left[
(1+x_0^{-1})^4 + \delta\right]
\label{77}\ee
Here we have also used
\be \lim_{\epsilon\rightarrow 0} x_1 =0\label{78}\ee
to resolve the ambiguity in sign.

For the expansion we require
\be
{\epsilon\over {\vert 1+x\vert}} \ll 1\label{79}\ee
which leads to a limit for the opening half-angle of the beam 
$\theta=-\pi +\arccos\mu_c $
\be
\epsilon\simeq \theta^2 \ll \delta^{1/4} \tan^{-{1\over 2}}\phi= \delta^{1/2}\quad{\rm for}\ 
\tan\phi=\delta^{-1/2}\label{80}\ee
The limiting beam opening half-angle is about 10 arcminutes for $\delta=10^{-10}$. For all $\epsilon$ satisfying Eq.\ref{80} the parallel
velocities $v_\parallel$ are always larger than the phase velocities in the
range where instability occurs.

In Fig.\ref{correct_a} we show the phase velocity modification, $V\Re x_1$,
for different parameters. The phase velocity modification scales
approximately linearly with $\epsilon$, unless $a_{\rm R}$ 
is close to $a_{\rm min}$, which reflects the dominance of the first order
term in $x_1$ over the second order term in Eq.\ref{75} for all $a_{\rm R}$
not too close to $a_{\rm min}$. Fig.\ref{correct_y} shows the 
growth rate modification, $\gamma_1$, for different parameters. 

In Fig.\ref{corr_growth} we show the modified growth rate, $\gamma+\gamma_1$, 
as a function of the modified phase velocity, $a_{\rm R}+V\Re x_1$. 
To be noted from the figure are 
\begin{itemize}
\item The growth rate increases if the beam has a finite opening angle.
\item The phase velocity range where instability occurs is only marginally
extended to phase velocities smaller than $a_{\rm min}$.
\end{itemize}

\section{Discussion}
We have calculated the excitation of longitudinal, electrostatic turbulence
subsequent upon the injection of a relativistic electron-proton beam into
a cold background plasma. In contrast to electromagnetic waves,
which scatter the particles in pitch angle but preserve their kinetic energy
until the distribution is isotropized,
the electrostatic waves change the particles' energy until a 
plateau distribution is established.
Our results can be applied to models of AGN and GRB,
which are based on a relativistic blast waves traversing the 
ambient medium. 

As one example of such models, 
two of us have recently published a calculation of an electromagnetic
two-stream instability in a channeled relativistic outflow, a jet, and shown
that the isotropization of the incoming interstellar particles would 
provide relativistic particles in the AGN jets, whose high energy emission
has characteristics typical of BL Lacertae objects
(Pohl \& Schlickeiser \cite{ps00}). The calculations presented here expand
on these results. We have shown that for parameters suitable to explain
the high energy emission spectra of AGN, the electrostatic instability
is much faster than the electromagnetic instability, even in case of a 
weak dispersion of the incoming beam. The beam will therefore relax to
a plateau-distribution in $p_\parallel$,
which is, however, still unstable with respect to electromagnetic waves. 
The spectrum of electromagnetic waves excited by a plateau-distribution 
in $p_\parallel$ will not be a $k^{-2}$ spectrum as in the case of a
cold beam, for it can be understood as a superposition of the wave spectra
produced by cold beams with different energies, each of which would be
a $k^{-2}$ spectrum with energy-dependent upper and lower limits in $k$ 
Nevertheless, 
isotropization in the collimated blast wave would still occur on
roughly the same time scale as calculated by Pohl \& Schlickeiser (\cite{ps00}).

\begin{figure}
\resizebox{\hsize}{!}{\includegraphics{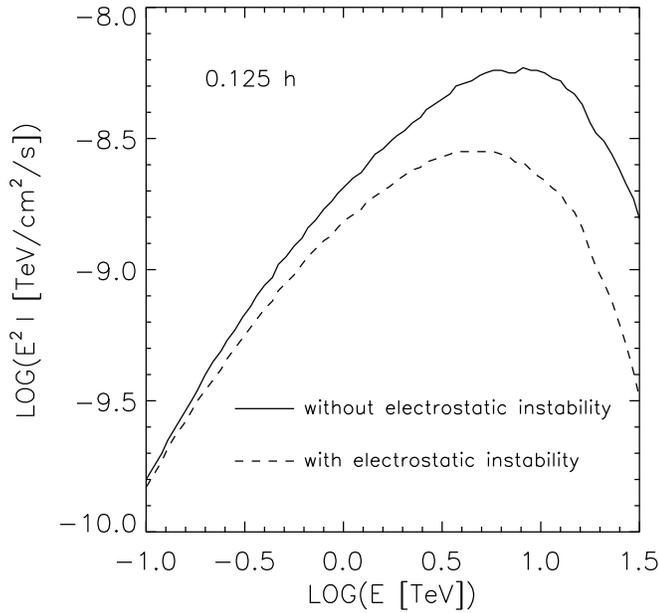}}
\caption{$\pi^0$-decay spectra calculated with and without the
electrostatic instability. 
The spectra refer to a observed time of 0.125 hours after
the blast wave ran over an isolated cloud.
At this time the spectrum of swept-up particles has 
evolved very little.
At the peak, about 50\% of the beam particle energy is lost due to the 
effect of the
electrostatic instability. There is little effect at smaller photon energies.}
\label{pinull}
\end{figure}

\begin{figure}
\resizebox{\hsize}{!}{\includegraphics{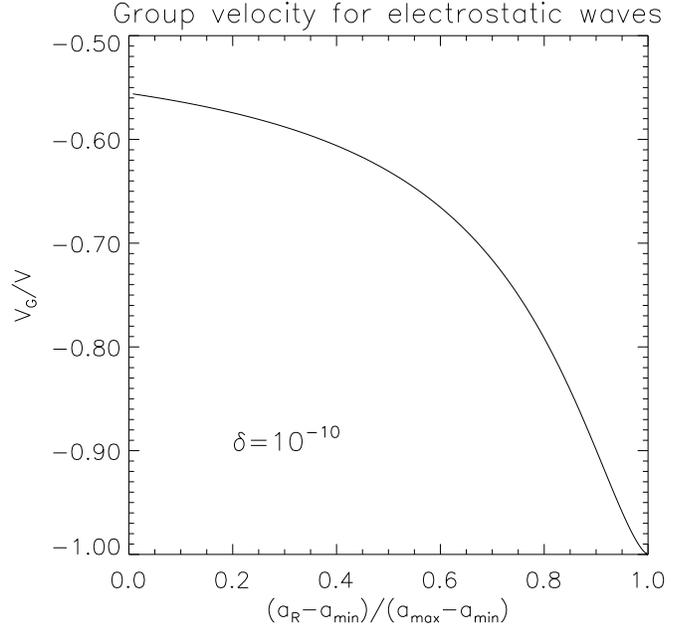}}
\caption{The group velocity of electrostatic waves as a function of phase velocity for the initial ideal beam. The group velocity is always between
about 50\% and 100\% of the beam velocity, which result is fairly independent
of the weighted beam intensity, $\delta$.}
\label{vgroup}
\end{figure}

Therefore, the main effect of the electrostatic instability is a
change in the spectrum of swept-up particles.
Instead of the prior result for the differential sweep-up rate
in the absence of the electrostatic instability
\be
\dot N(\gamma)\bigg\vert_{\rm no\ e.s.i.} = N_0\ \delta (\gamma-\Gamma)\ 
\label{noesi}\ee
we now obtain 
\be
\dot N(\gamma)\bigg\vert_{\rm e.s.i.} = {{N_0}
\over {\Gamma - 1}}\ \Theta (\Gamma-\gamma) \label{esi}\ee
In the radiation spectra this change corresponds to a reduction of
efficiency by roughly 50\% at the highest energies and little
change at smaller energies, because in the inelastic pp collisions the pion source power scales roughly with the kinetic energy of the incident proton,
hence the pion spectra are dominated by the pions produced by the protons
of high energy for flat proton spectra.
In Fig.\ref{pinull} we show the spectra of $\pi^0$-decay \grs for
the case of a fast electrostatic instability (Eq.\ref{esi})
in comparison with the case of the electromagnetic instability only
(Eq.\ref{noesi}). At photon energies well below the peak energy of
the $\nu F_\nu$-spectra the difference in spectral index is only 
$\Delta s\simeq 0.1$. Similarly, small changes can be expected for the injection
spectra of secondary electrons and hence for the synchrotron spectra 
at optical to X-ray frequencies.
Therefore the isotropization and radiation conclusions
drawn by Pohl \& Schlickeiser (\cite{ps00}) are still valid.
About half of the energy flux carried by the incoming beam is channeled
into electrostatic turbulence. It is instructive to calculate the group
velocity of the waves with which the wave energy is transported. In 
Fig.~\ref{vgroup} we show the group velocity for the initial ideal beam
(Eq.\ref{28}) in the phase velocity range where instability occurs. The group velocity is fairly independent of the weighted beam intensity, $\delta$, and it 
is always larger than $0.5\,V\simeq 0.5\,c$. Therefore, 
the energy carried by the electrostatic waves is rapidly transported 
toward the backside of the blast wave.
Because the waves carry 50\% of the energy flux of the incoming beam with a 
transport velocity of 50\%--100\% of the beam velocity, 
the energy densities in the final wave
field and the initial beam should be similar.
\be
U_{\rm W} = \int da_{\rm R}\ I(a_{\rm R}) \simeq n_{\rm i}\,m_p c^2\,\Gamma
=n_{\rm i}^\ast\,m_p c^2\,\Gamma^2
\label{81}\ee
The high energy density of waves has two important aspects, which are 
briefly discussed in the following sections.
\subsection{Nonlinear evolution of the wave field}
Quasi-linear theory is not valid under all physical conditions. When the 
energy density of the wave field is not small, nonlinear effects may become
important.

One important effect is nonlinear Landau damping (Sagdeev \& Galeev
\cite{sg69}). This is the interaction between thermal particles and the beat of
two electrostatic waves: if the phase velocity of the beat wave is almost
equal to the thermal velocity of charged particles in the background plasma, 
the particles would Landau-damp the beat wave
and hence be heated. If the remaining wave has a superluminal phase velocity, 
it will not resonate with the beam particles anymore. As a consequence, 
the number of scattering processes between the beam particles and the waves, 
which are responsible for the production
of a plateau in the beam distribution function,
would be drastically reduced. This process reduces the wave intensity in the 
resonant phase velocity range as long as the intensity of waves with
superluminal phase velocity is not so strong that nonlinear interactions in
this regime cause wave energy to be retransferred to resonant subluminal
phase velocities.

The nonlinear Landau damping is a
process resulting from weak turbulence theory: although waves interact, the
wave energy is small enough that individual waves are identifiable.
If the wave energy is too high, strong turbulence theory applies (Zakharov
\cite{za71}). The charge distribution in the background plasma, which results
from the electric field of the waves (essentially the RHS of Eq.\ref{12}),
becomes so strong that it affects the dispersion relation. As a result, 
the wave field will collapse into cavitons (Goldman \cite{go84}) in which 
the plasma has a low density. This effect will depend on the temperature 
of the background plasma, and on whether or not the diffusive particle 
motion can effectively smear out the depleted cavitons, 
and thus impede their build-up.

Without having performed a detailed study of these effects in a background medium 
with possibly time-variable temperature caused by heating processes,
we can only speculate on the impact of the nonlinear effects, which would 
increase the time scale for the plateau production of the beam (Eq.\ref{65}).
However, because that time scale is more than two orders of magnitude 
shorter than the time scale for the electromagnetic instability, even if important such nonlinear effects would likely still permit the production of a 
plateau on a timescale much faster than the 
isotropization.
 
\subsection{Where does the wave energy go?}
Part of the wave energy will be transferred to the background plasma in 
the blast wave. We have already discussed the nonlinear Landau damping, 
in which thermal particles would absorb the beat of two electrostatic waves. 
But, in the general case, the natural width of the resonance Lorentz
profile combined with the Maxwellian distribution function allows thermal
particles to Landau damp the waves with $v_\phi \simeq c$, albeit with a small
probability. 

We do not know the fraction of the wave energy density that is channeled into
heating processes of the blast wave plasma. Here we concentrate on a 
qualitative discussion of the possible effects.

The wave energy is transported quickly toward the backside of the collimated
blast wave. At the transition between the blast wave and its wake, the 
dispersive properties of the blast wave change. We can, therefore,
expect that a part of the wave field is reflected at the backside, and possibly 
again reflected at the front side and so forth. If the reflection coefficient
is not small, the wave energy density inside the blast wave would increased,
until nonlinear effects at the front and backside cause the waves to escape. 
The heating rate would be similarly enhanced.

What would be the effect of strong heating in the blast wave? As an 
extreme example consider that the full wave energy flux is available
for homogeneously heating the blast wave.
The heating rate then is
\be
\dot T \simeq {{(\gamma_{\rm a} -1) n_{\rm i}^\ast \,m_{\rm p}c^3\,\Gamma^2}
\over {2 n_{\rm b} d\,k}}\simeq
10^6\, {{n_{\rm i}^\ast \,\Gamma_2^2}\over {n_8\,d_{13}}}\quad
{\rm K/sec}\label{82}\ee
where we have used the scaled variables $n_8= 10^{-8} n_b$,
$d_{13}=10^{-13} d$, and $\Gamma_2 =10^{-2} \Gamma$. This heating rate would be 
too large to be balanced by any cooling process of a non-relativistic
plasma. Therefore run-away heating would occur until non-thermal cooling processes set in at semi-relativistic energies.

The existence of radio galaxies implies that the collimation of AGN jets 
is well maintained on linear scales of hundreds of kiloparsecs. A
semi-relativistic plasma jet could hardly remain collimated
on these scales, because the expansion by internal pressure alone would
cause a substantial opening of the jet.
We may take the observed collimation of jets in radio galaxies as empirical
evidence against excessive heating. For the fate of the 
electrostatic wave field, this observation indicates that the reflection
efficiency of waves at the backside of the blast wave is small, whatever
the cause. 

Since the waves carry momentum, even a small reflection would transfer momentum
to the background plasma at the backside of the blast wave, which would then
"boil off" the blast wave and fill its wake. The waves that have escaped
the blast wave would fill the wake and the regions around it, and be finally 
damped by the interstellar plasma in the host galaxy of the AGN, thus giving
rise to additional heating of the interstellar medium around the AGN jet.

\begin{acknowledgements}
Partial support by the Verbundforschung, grant {\it DESY-05AG9PCA}, is gratefully acknowledged.
\end{acknowledgements}

\end{document}